\documentclass[twoside,journey]{IEEEtran}
\usepackage{makecell}%1
\usepackage{array}%1
\usepackage{graphicx,amssymb,amsmath}
\usepackage{multicol}
\usepackage[noadjust]{cite}
\usepackage{caption}
\usepackage{setspace}
\usepackage{subfigure}%1
\usepackage{graphicx}
\usepackage{float}%1
\usepackage {url}
\usepackage{stfloats}
\usepackage{amsthm,pifont}
\usepackage{cases,subeqnarray}
\usepackage{bm,multirow,bigstrut}
\usepackage{amsmath, amsthm, amssymb}
\usepackage{textcomp}
\usepackage{latexsym,bm}
\usepackage{booktabs}
\usepackage{xcolor}
\usepackage{mathtools}
\usepackage{dsfont}
\usepackage{extarrows}
\usepackage{epsfig}
\usepackage{epsfig}
\usepackage{epstopdf}
\usepackage[noend]{algpseudocode}
\usepackage{algorithmicx,algorithm}

\usepackage{cite}
\usepackage{bm}
\usepackage{cleveref}
\usepackage{multicol}       % table
\usepackage{multirow}       % table
\usepackage{array}          % table
\usepackage{colortbl}
\usepackage{makecell}
\definecolor{crimson}{RGB}{192,0,0}         % color crimson
\definecolor{navy}{RGB}{47,85,151}         % color crimson
\usepackage{bbding}
\usepackage{graphicx}
\usepackage{booktabs}

\usepackage{colortbl}

\theoremstyle{plain}
\theoremstyle{remark}

\IEEEoverridecommandlockouts

\begin{document}
%----------------------------title&author&thanks----------------------------
\title{Energy-Efficient RIS-Aided Cell-Free Massive MIMO Systems: Application, Opportunities, and Challenges}

\author{Yu Lu, Jiayi~Zhang,~\IEEEmembership{Senior Member,~IEEE}, Enyu Shi, Peng Zhang, Derrick Wing Kwan Ng,~\IEEEmembership{Fellow,~IEEE},\\ Dusit Niyato,~\IEEEmembership{Fellow,~IEEE}, and Bo Ai,~\IEEEmembership{Fellow,~IEEE} 
\thanks{Y.~Lu,  J. Zhang, E. Shi, P. Zhang, and B. Ai are with the School of Electronic and Information Engineering, Beijing Jiaotong University, Beijing 100044, China. (email: jiayizhang@bjtu.edu.cn).}
\thanks{D. W. K. Ng is with the School of Electrical Engineering and Telecommunications, University of New South Wales, Sydney, NSW 2052, Australia. (e-mail: w.k.ng@unsw.edu.au).}
\thanks{D. Niyato is with the College of Computing and Data Science, Nanyang Technological University, Singapore (e-mail: dniyato@ntu.edu.sg).}
\thanks{This work was supported in part by National Natural Science Foundation of China under Grant 62471027 \& 62221001, in part by the Fundamental Research Funds for the Central Universities under Grant 2022JBQY004, and in part by ZTE Industry-University-Institute Cooperation Funds under Grant No. IA20240319002. The work of D.Niyato was supported by the National Research Foundation, Singapore, and Infocomm Media Development Authority under its Future Communications Research \& Development Programme, Defence Science Organisation (DSO) National Laboratories under the AI Singapore Programme (FCP-NTU-RG-2022-010 and FCP-ASTAR-TG-2022-003), Singapore Ministry of Education (MOE) Tier 1 (RG87/22), the NTU Centre for Computational Technologies in Finance (NTU-CCTF), and Seitee Pte Ltd.}
}
\maketitle
\vspace{-1.5cm}
%----------------------------abstract----------------------------

\begin{abstract}
Reconfigurable intelligent surfaces (RIS)-assisted cell-free massive multiple-input multiple-output (CF mMIMO) systems have emerged as a promising technology for sixth-generation communication systems. These systems capitalize on RIS to minimize power consumption, thereby achieving consistent performance and enhancing communication quality through the establishment and shaping of auxiliary signal propagation pathways between access points (APs) and users. However, integrating RIS into existing CF mMIMO infrastructures presents several technical challenges. This study delves into the signal transmission scheme and deployment architecture of RIS-aided CF mMIMO systems, addressing inherent challenges such as interference induced by RIS and the increased complexity in beam alignment. Furthermore, we address the complexities arising from the joint optimization of the reflection phase of RIS and beamforming technology at the APs, intending to fully exploit the reflection capabilities of RISs and beamforming technology to maximize the energy efficiency (EE) of the system. To overcome these challenges, we propose cooperation communication to suppress RIS-induced interference, beam tracking, and joint optimization to improve system EE. We also present specific examples of cooperative communication under the constraint of electromagnetic interference and the beam tracking of a mobile system. Additionally, we emphasize important research directions for RIS-aided CF mMIMO systems, aiming to inspire future investigations.
\end{abstract}
%----------------------------keywords----------------------------

%\newpage
\IEEEpeerreviewmaketitle
\section{Introduction}
In sixth-generation (6G) wireless systems, enhancing system capacity is crucial. Cellular systems face persistent inter-cell interference,  especially for users near cell boundaries. Cell-free (CF) networking has been proposed to address this. Unlike cell-centric networks, CF systems support user-centric transmissions, where clusters of access points (APs) jointly serve devices without strict cell boundaries. This cooperation among APs effectively mitigates inter-cell interference, significantly improving network capacity \cite{shi2024ris,peng2023resource1}.

\begin{figure*}[t]
\centering
\includegraphics[scale=0.24]{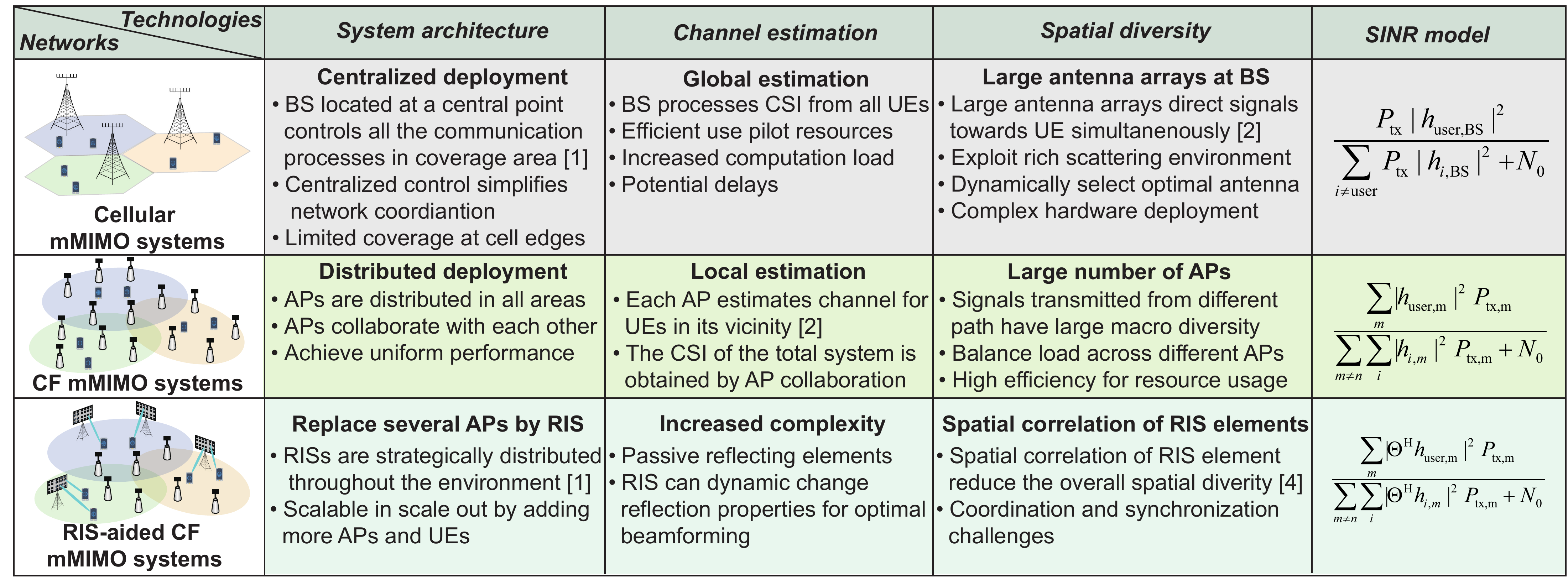}
\caption{{The major differences in cellular, CF mMIMO, and RIS-aided CF mMIMO systems.}\label{CF_compare}}
\end{figure*}

{The 6G networks aim for a 100× increase in energy efficiency (EE) over 5G networks. However, the extensive deployment of APs in cell-free massive multiple-input multiple-output (CF mMIMO) systems poses a significant energy consumption challenge  \cite{peng2023resource1}. Reconfigurable intelligent surfaces (RIS) can help by replacing some APs and reflecting signals to users, especially those at the cell edges, to improve EE without compromising performance. Fig. \ref{CF_compare} compares cellular mMIMO systems, CF mMIMO systems, and RIS-aided CF mMIMO systems. In the CF system, local channel estimation necessitates collaboration among APs. Moreover, CF networks empower user equipment (UE) to exploit spatial diversity, enhancing system capacity beyond what traditional cellular systems can achieve, which is often hindered by cell partitioning \cite{zhang2021beyond}. Integrating RIS in the CF systems boosts signal transmission, and enhances spatial diversity and system capacity, overcoming the limitations brought by the massive deployment of APs for broader spatial coverage and higher capacity \cite{pan2021reconfigurable}. }

\begin{figure*}[t]
\centering
\includegraphics[scale=0.25]{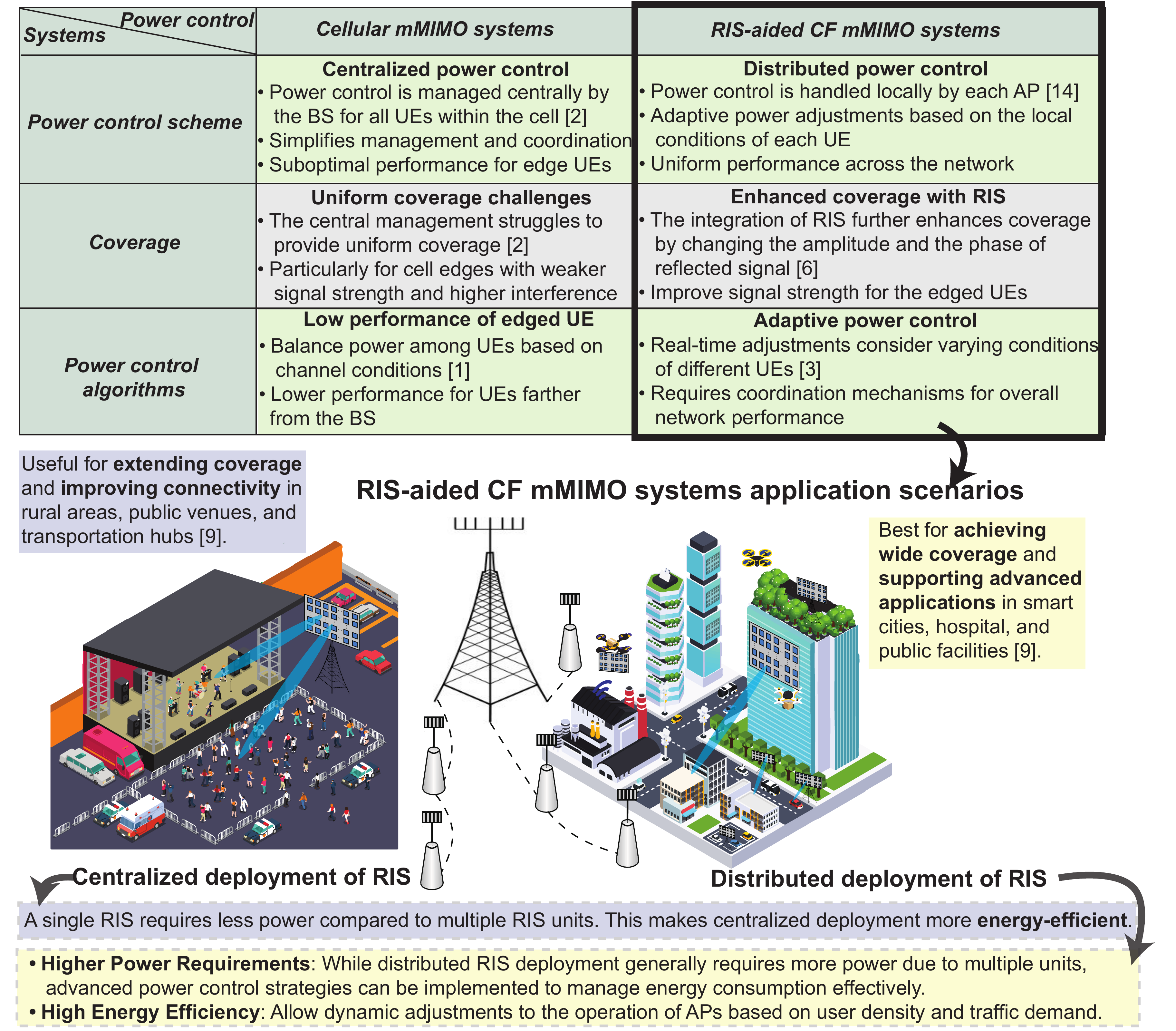}
\caption{{The major differences in power control for cellular systems and RIS-aided CF mMIMO systems and the centralized and distributed deployments of RIS.}\label{compare_scene}}
\end{figure*}

{In RIS-aided CF mMIMO systems, spectral efficiency (SE) and EE are important performance indicators in the RIS-aided CF mMIMO system. In particular, SE is enhanced through optimized beamforming via adaptive RIS reflection, while EE improvement stems from energy-saving design strategies.} Regarding power control, as depicted in Fig. \ref{compare_scene}, CF mMIMO systems outperform cellular mMIMO systems in dynamic and localized power adjustments due to their distributed nature. Integrating RIS into CF mMIMO systems reduces power consumption and refines the transmission environment, enhancing signal strength and mitigating interference. {For instance, \cite{jin2023energy} and \cite{lyu2023energy} demonstrated improved system EE through power allocation, and the joint optimization of beamforming at the APs and phase shifts at the RIS.} {These studies indicate that despite the extra signaling overhead from phase shift adjustments and channel estimation brought by RIS integration, proper design and algorithmic advancements in strategic optimization and RIS deployment help to achieve an effective balance between the overhead and performance improvements.} 

{However, existing studies including \cite{jin2023energy} and \cite{lyu2023energy} often underplay the difficulties that arise from integrating RIS into CF mMIMO systems. These include increased mutual coupling among closely packed RIS elements, which can interfere with their reflective properties and degrade overall performance. Additionally, RIS can amplify electromagnetic interference (EMI) and inter-operator interference, particularly through unintended signal reflections that complicate beamforming and increase disruptions to other wireless systems. These issues are inherent to RIS-aided networks and are further intensified by the distributed and cooperative framework of CF mMIMO systems.}

{Distinguishing from \cite{shi2024ris} that encompasses a wide-ranging examination of system architecture and application scenarios, our study emphasizes improving EE within RIS-aided CF mMIMO systems. Our key contribution lies in the strategic enhancement of EE, achieved through the refinement of power control mechanisms.} The main contributions of our work are summarized as follows:
%\vspace{-3mm}
\begin{itemize}
\item
This article thoroughly explores the challenges linked to integrating RIS into existing CF mMIMO systems, particularly with a focus on EE. Our examination commences with a comprehensive overview of the signal transmission process within the RIS-aided CF mMIMO system, highlighting distinctions between the signal transmission mechanisms of cellular systems, CF mMIMO systems, and RIS-aided CF mMIMO systems. Additionally, we delve into the practical deployment of the RIS in the existing system, taking into account specific characteristics and requirements of scenarios.
\item
Moreover, we investigate the energy consumption challenges arising from incorporating RIS into CF mMIMO systems. These challenges encompass elevated power usage attributed to interference induced by RIS, such as mutual coupling, EMI, and inter-operator interference. Additionally, there is increased complexity in beam alignment and the joint optimization of beamforming at the APs and phase shifts at the RIS to maximize EE. %To address these challenges, we propose potential solutions, including cooperative communication, beam tracking techniques, and optimization coupling methods.
\item
Moreover, we present a practical framework for a cooperative communication-based RIS-aided CF mMIMO system, considering interactions among the APs, the RIS, and users to tackle the interference induced by RIS. Within the framework, we introduce beam tracking methods tailored for various practical scenarios and propose efficient joint optimization techniques for beamforming at APs and reflection at RIS. Lastly, we outline future research directions for further investigation. 
\end{itemize}

%\newcounter{mytempeqncnt}
\section{Signal transmission scheme and system deployment}
%\vspace{-3mm}
\vspace{-1mm}
In this section, we illustrate the signal transmission schemes in a cellular mMIMO system, a CF mMIMO system, and an RIS-aided CF mMIMO system. We then explore the practical deployment of the RIS-aided CF mMIMO system, including both centralized and distributed RIS deployment strategies. %, along with in-depth insights into typical application scenarios. 
\vspace{-3mm}
\subsection{Signal Transmission Scheme}

\begin{figure*}[!t]
\centering
\includegraphics[scale=0.19]{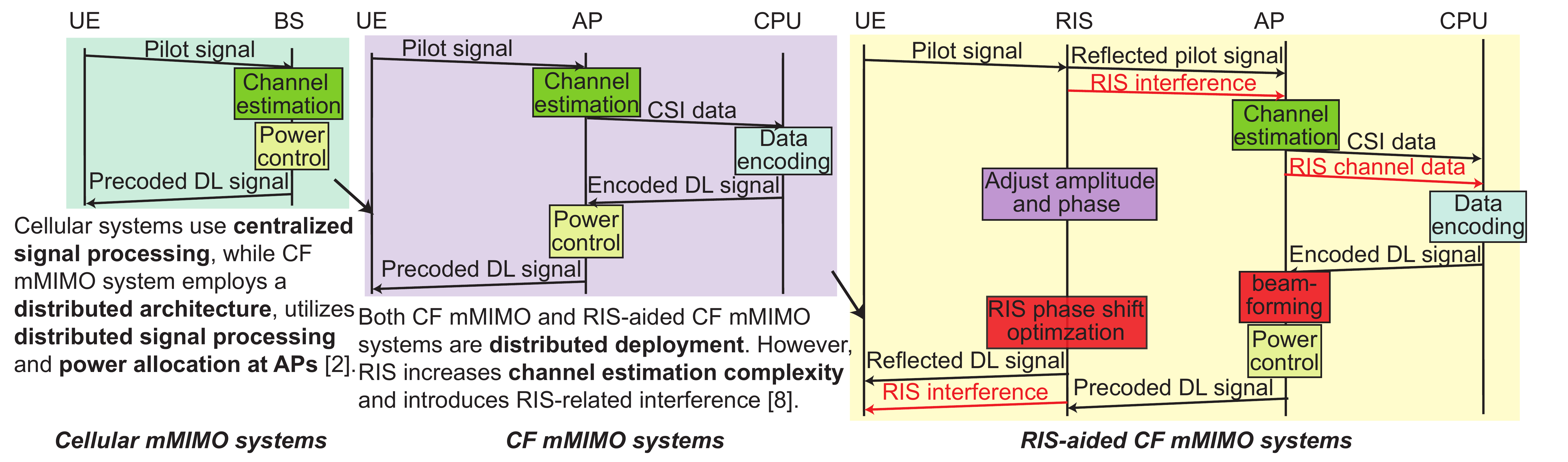}
\caption{{Signal transmission procedure of a cellular mMIMO system, a CF mMIMO system, and an RIS-aided CF mMIMO system.}\label{power_scheme}}
\end{figure*}

{Fig. \ref{power_scheme} delineates the signal transmission processes inherent in a cellular mMIMO system, a CF mMIMO system, and an RIS-aided CF mMIMO system. This includes the uplink channel estimation and downlink signal transmission phases. In the cellular mMIMO system, the BS is responsible for CSI acquisition through pilot signals from UEs and then proceeds with signal transmission based on these estimates. Besides, power control is centralized at the BS to manage transmission power across cells \cite{zhang2021beyond}. While effective, this centralized approach can lead to inefficiencies and limitations in dynamic environments. In contrast, the distributed nature of the CF mMIMO system offers a more adaptable framework for uplink and downlink signal transmission. Each AP in the CF mMIMO system estimates the channels locally in uplink transmission, and these local estimates are then aggregated at the CPU to establish a comprehensive view of the UE data \cite{peng2023resource1}. This distributed channel estimation approach allows quicker responses to local channel conditions, reducing latency and enhancing the system's ability to adapt to changes. For downlink operations, the CPU performs encoding while delegating power allocation to the APs, which can adapt their transmit power based on local conditions. This decentralization of power control not only reduces the computational load at the CPU but also increases the system's flexibility and scalability, thereby improving adaptability in dynamic wireless environments \cite{zhang2021beyond}. }

{In an RIS-aided CF mMIMO system, the RIS serves as a substitute for APs, effectively reducing the system's AP count and energy consumption. It is important to note that the RIS, depicted in Fig. \ref{power_scheme}, is a passive component rather than an energy-harvesting device that relies on energy supplied by the APs for its functioning \cite{albanese2024ares}. The design of RIS phase shifts is strategically based on statistical CSI, which is less time-varying than the instantaneous CSI. This approach allows for less frequent updates of the RIS phase shifts, thereby significantly reducing the overhead associated with channel estimation \cite{zhi2022two}. The complexity of CSI acquisition rises with the growing number of antennas and channels. To enhance the process, methods such as pilot-based estimation and AI-driven machine learning can be utilized to achieve greater system performance. Furthermore, in both CF mMIMO and RIS-aided CF mMIMO systems, distributed power control at APs is optimized for energy efficiency and adaptability to varying conditions through long-term and short-term strategies. Machine learning and AI can also be utilized to improve power control dynamics. The main difference lies in their signal transmission schemes, i.e., direct transmission in CF mMIMO versus RIS reflection transmission in RIS-aided systems \cite{jin2023energy}.}

\vspace{-3mm}
\subsection{Centralized and Distributed Deployment of RIS}
The deployment of reflecting elements in RIS-aided CF systems can be implemented using two strategies: centralized and distributed deployments (as depicted in Fig. \ref{compare_scene}). Centralized deployment consolidates elements on a single RIS, simplifying power allocation and reducing computational complexity and pilot overhead, with centrally optimized beamforming for better signal strength and coverage. Distributed deployment spreads elements across multiple RIS units, complicating channel estimation but enhancing spatial diversity and resistance to fading. Each RIS unit optimizes beamforming locally while coordinating with a central controller, improving coverage and connectivity. {In actual deployment scenarios, distributed configurations typically outperform centralized ones in various aspects. Despite higher power requirements, the distributed setups provide improved independence and space-time diversity with an equivalent number of RIS elements, making them more adaptable to the specific needs of the communication system \cite{ma2023cooperative}.} In addition, hybrid deployment which combines centralized and distributed strategies is practically applicable. This approach places a central RIS offering broad coverage, while smaller distributed RIS units provide targeted enhancements. This approach balances the simplicity of centralized control with the performance benefits of distributed setups, optimizing coverage, and signal quality, and managing complexity and power consumption effectively.

In summary, cellular mMIMO systems rely on centralized power control by the BS, which can lead to performance degradation at the edges. CF mMIMO systems implement distributed power control at APs, ensuring uniform performance and enhancing system flexibility. RIS-aided CF mMIMO systems combine distributed AP control with passive RIS control, optimizing power allocation and improving system adaptability. Additionally, reflecting elements in RIS can be deployed in centralized, distributed, or hybrid configurations, each offering unique advantages in terms of power management, spatial diversity, and overall system efficiency.

\section{Challenges and Solutions of RIS-aided CF mMIMO implementation}

In this section, we discuss the three key challenges associated with implementing the RIS-aided CF mMIMO system.
\vspace{-1em}
\subsection{RIS-induced Interference}
The primary challenge faced by existing CF mMIMO systems is the high energy consumption due to extensive AP deployment. To mitigate this, we propose substituting multiple APs with RIS to reduce overall energy consumption. However, RIS integration introduces interference, exacerbating energy consumption issues as interference energy is associated with transmitted signals. Specifically, mutual coupling effects and EMI during signal transmission, stemming from interference within RIS elements, worsen with spatial correlation between RIS elements. Additionally, RIS integration intensifies inter-operator interference, directly proportional to transmitted signal power. While these challenges are prevalent in other RIS-aided wireless networks, CF mMIMO systems, employing large-scale fading decoding (LSFD) at the APs \cite{peng2023resource1}, are less prone to interference but require precise power control. Inconsistent power management among densely deployed APs can degrade system performance. Effective interference mitigation demands varied cooperation strategies among APs, setting RIS-aided CF mMIMO apart from other RIS-aided networks. Subsequent sections explore these adverse effects.

\subsubsection{Mutual Coupling}
{Mutual coupling in RIS-aided CF mMIMO systems is exacerbated by the proximity of RIS elements, which can lead to signal integrity degradation due to crosstalk and interference. This negative effect not only hampers the individual elements' ability to autonomously adapt their reflection properties, compromising overall performance but also results in increased power consumption as the system works to counteract the interference. The need for additional signal processing techniques and control mechanisms to mitigate these impacts can further diminish the system EE. Furthermore, in dense RIS deployments or larger RIS sizes, ignoring mutual coupling can significantly impair channel estimation. Even with mutual coupling awareness, tight RIS element spacing can still degrade performance \cite{qian2021mutual}.}

\subsubsection{Electromagnetic Interference on RIS}
Stemming from both man-made devices and natural sources of radiation, EMI poses significant challenges for RIS systems. EMI reaches the RIS from a variety of spatial directions before being re-radiated. In practice, electromagnetic radiation can propagate through free space and interfere with the communication signals of other wireless systems or devices. Such interference can lead to degradation in system performance. For example, re-radiated EMI from the RIS in unintended directions can disrupt the intended signal transmission. Moreover, this interference can spread through the RIS structure to affect other components of the system, potentially disrupting their operations. EMI might also propagate through the RIS's connected cables or shared power lines, further jeopardizing system functionality \cite{shi2023uplink}. 
\subsubsection{Inter-operator Interference}
In practice, operators use different frequency bands to reduce interference, but inter-operator interference can still occur due to frequency leakage, leading to performance degradation and lower quality of service. Beamforming intended for one operator may unintentionally interfere with adjacent operators \cite{shi2024ris}. The integration of RIS can exacerbate this interference through signal reflections, causing unintended beamforming effects and increasing interference with other operators' signals. Variations in channel conditions and RIS configuration dynamically affect interference levels, making it harder to manage signal quality. Mitigating this interference requires additional energy and reduces power allocation flexibility.

\subsection{Beam Alignment}
{In the RIS-aided CF mMIMO system, beam alignment is intricately challenged by the need for precise coordination among APs and the RIS. Unlike traditional cellular networks where each cell handles beam alignment in isolation, the CF mMIMO architecture relies on a collaborative approach among APs to direct coherent signals toward users \cite{zhang2021joint,ma2023cooperative}. This collaboration is essential for enhancing spatial multiplexing and reducing inter-cell interference, which is a typical issue in cellular systems with limited inter-cell coordination.}
\\
{Functioning as a passive device, the RIS cannot generate signals but excels in reflecting them. Consequently, when integrated into the CF mMIMO system, the RIS replaces a subset of APs by reflecting the target signal, enhancing the accuracy of channel estimation through its involvement in pilot signal transmission. However, incorporating RIS into the CF mMIMO system inherently complicates the transmission environment. The RIS requires the APs to estimate the cascaded channels it introduces, which increases the computational burden for channel estimation. Furthermore, the integration of RIS in CF mMIMO systems, with their distributed APs, can heighten synchronization challenges, impacting time and frequency domains. Mismatches in AP frequency synthesizers can cause beam misalignments and heightened interference, critical for coherent signal combining at user terminals \cite{shi2024ris}. To mitigate these issues, distributed synchronization protocols are crucial, which enable the exchange of timing and phase information across the network, ensuring system-wide synchronization and proper alignment. This synchronization is key to optimizing the performance and reliability of RIS-aided CF mMIMO systems.}
\vspace{-1em} 
\subsection{Joint Optimization}
Optimizing both the RIS reflection phase and AP beamforming is crucial in RIS-aided wireless communication systems to enhance signal strength and system EE by directing beams accurately to target users. However, in CF mMIMO systems, coordinating beamforming at APs and phase design at RIS presents a unique challenge. The dense AP deployment increases complexity, requiring coordination among distributed APs, resulting in higher computational complexity and signaling overhead compared with other RIS-aided communication systems. Thus, developing advanced joint optimization strategies for RIS-aided CF mMIMO systems is essential yet challenging \cite{zhang2021beyond}.

In summary, implementing RIS-aided CF mMIMO systems faces three main challenges: RIS-induced interference, beam alignment, and joint optimization. RIS-induced interference complicates power management and affects signal quality. Beam alignment requires precise synchronization between APs and RIS to avoid misalignments. Joint optimization of RIS reflection and AP beamforming requires advanced strategies to manage computational and signaling overhead. Addressing these challenges is essential for enhancing the efficiency and performance of RIS-aided CF mMIMO systems.

\section{Promising solutions}
In this section, we discuss potential solutions in response to the issues explained in Section \uppercase\expandafter{\romannumeral3}.

\begin{figure*}[t]
\centering
\includegraphics[scale=0.28]{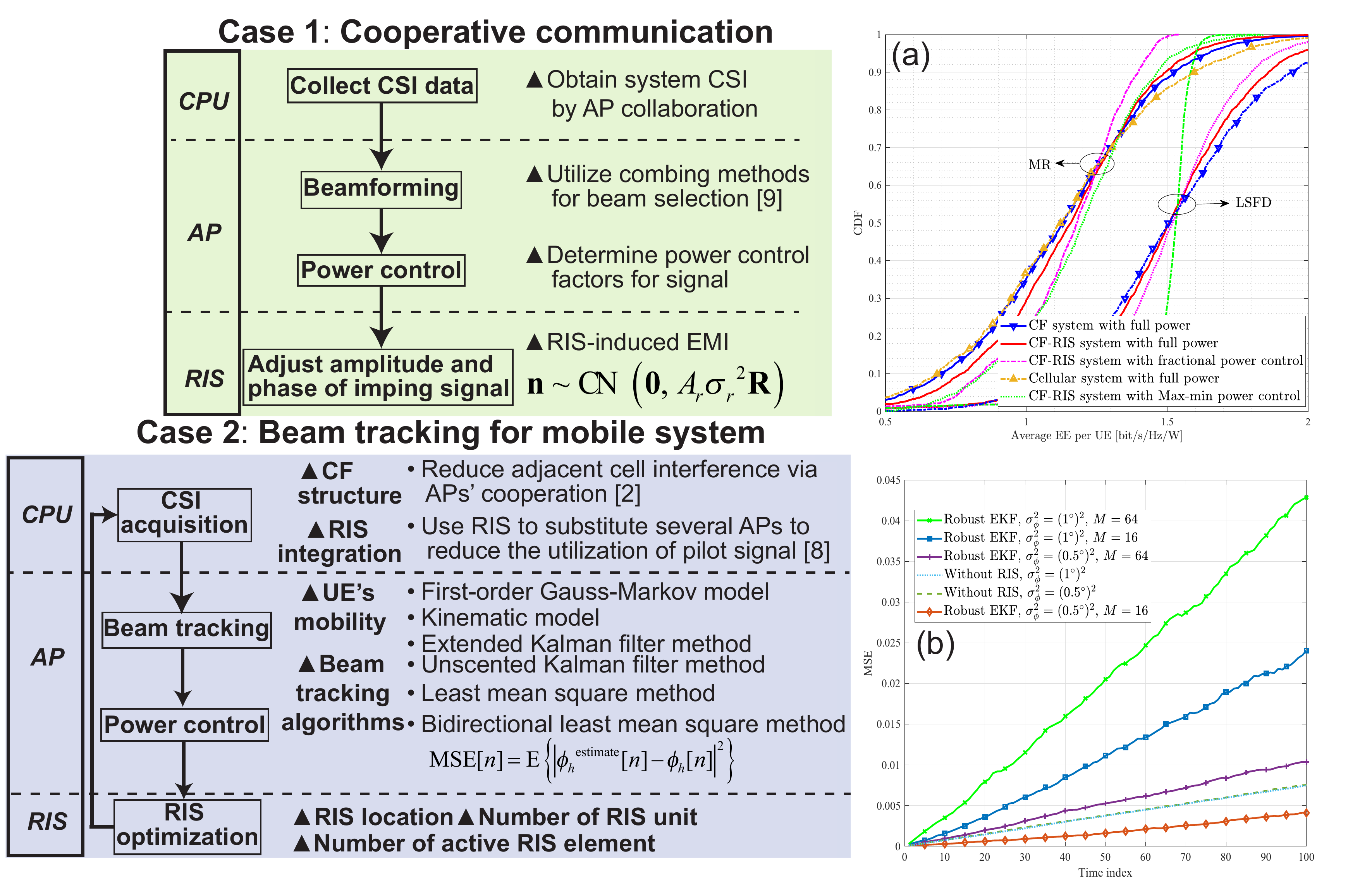}
\caption{{\textbf{Case 1}: Electromagnetic interference (EMI) restraint in an RIS-aided CF mMIMO system. EMI is quantified as additive interference in signal transmission. The fractional power control and max-min power control algorithms mitigate EMI's negative impacts, enhancing system performance. (a) shows performance under the maximum ratio (MR) combining and large-scale fading decoding (LSFD) methods at the APs \cite{shi2023uplink}. \textbf{Case 2:} Beam tracking in a time-varying RIS-aided CF mMIMO system with a mobile user while the AP and RIS remain static. A kinematic model characterizes user mobility, and a modified unscented extended Kalman filter (EKF) is used for precise channel beam tracking. (b) shows the mean square error under different covariance for the arrival of angle ($\sigma_{\phi}^2$) and RIS reflecting element ($M$) \cite{zhang2022ris}.}\label{case_study}}
\end{figure*}

\vspace{-1mm} 
\subsection{Cooperative Communication}
%\vspace{-2mm}
{Cooperative communication strategies for uplink systems with RIS address interference challenges like mutual coupling and EMI. In this distributed framework, APs, a CPU, and RISs collaborate to enhance communication efficiency and reliability. The CPU collects CSI to encode signals that are precisely tailored to current channel conditions and sends them to the APs. These APs then exploit beamforming techniques like maximum ratio (MR) combining to optimize signal reception. The LSFD method dynamically adjusts transmission power for efficiency. The RIS plays a critical role by reflecting signals to enhance strength and coverage without extra power consumption, addressing challenges like mutual coupling and EMI. This two-layer cooperation for interference elimination stands out for its focus on optimizing EE and managing the complexity introduced by RIS integration \cite{lyu2023energy,zhang2021joint}.}
{Furthermore, to balance the system overhead and performance in RIS-aided CF mMIMO systems, it is crucial to optimize resource allocation, design cooperative communication protocols, and integrate advanced signal processing techniques. This approach ensures efficient operation by balancing system demands with performance capabilities \cite{ma2023cooperative}.} {Different from \cite{qian2021mutual} which proposed an optimization algorithm that leverages the characteristics of RIS-induced interference, our method emphasizes system-specific optimization, focusing on enhancing signal strength and coverage while reducing interference introduced by RIS integration, making it particularly suitable for RIS-aided CF mMIMO systems.}

{In contrast to traditional cellular systems that often transmit at full power, leading to inefficient power utilization, our RIS-aided CF mMIMO system employs a fractional power control scheme. This assigns specific power control factors to each UE based on its channel condition, ensuring uniform service quality and consistent communication performance regardless of location or channel state. This approach is particularly effective in RIS-aided systems, where the integration of RIS can further achieve high-capacity communication coverage.}
\vspace{-1em}
\subsection{Beam Tracking}
{Beam tracking in RIS-aided CF mMIMO systems is essential for high-frequency bands, addressing the challenges of precise alignment due to narrow beams and blockage sensitivity. Our research presents an innovative beam tracking approach tailored to these systems, where RIS units are used in place of some APs to lower energy consumption without sacrificing performance. The passive RIS elements, which reflect but do not process signals, shift the computational load to the APs. To manage this, we've developed a beam tracking method that dynamically adjusts to the UE's movement and optimizes RIS element usage based on environmental changes. This method, which combines precise beam direction tracking with RIS element optimization, is distinctive in the field. We utilize advanced beamforming at the APs, including MR combining, for optimal signal reception. Coupled with algorithms like least mean square (LMS) and bidirectional LMS, these approaches dynamically adjust transmission power, ensuring accurate beam tracking in dynamic conditions \cite{zhang2022ris}.}

\vspace{-3mm}
\subsection{Optimization Coupling}

\begin{table*}[t]
\centering
\caption{{The illustration of the typical traditional methods and the machine learning method for joint optimizing both the RIS reflection phase and AP beamforming to maximize energy efficiency.}\label{opt}}
\vspace{-2mm}
\includegraphics[scale=0.25]{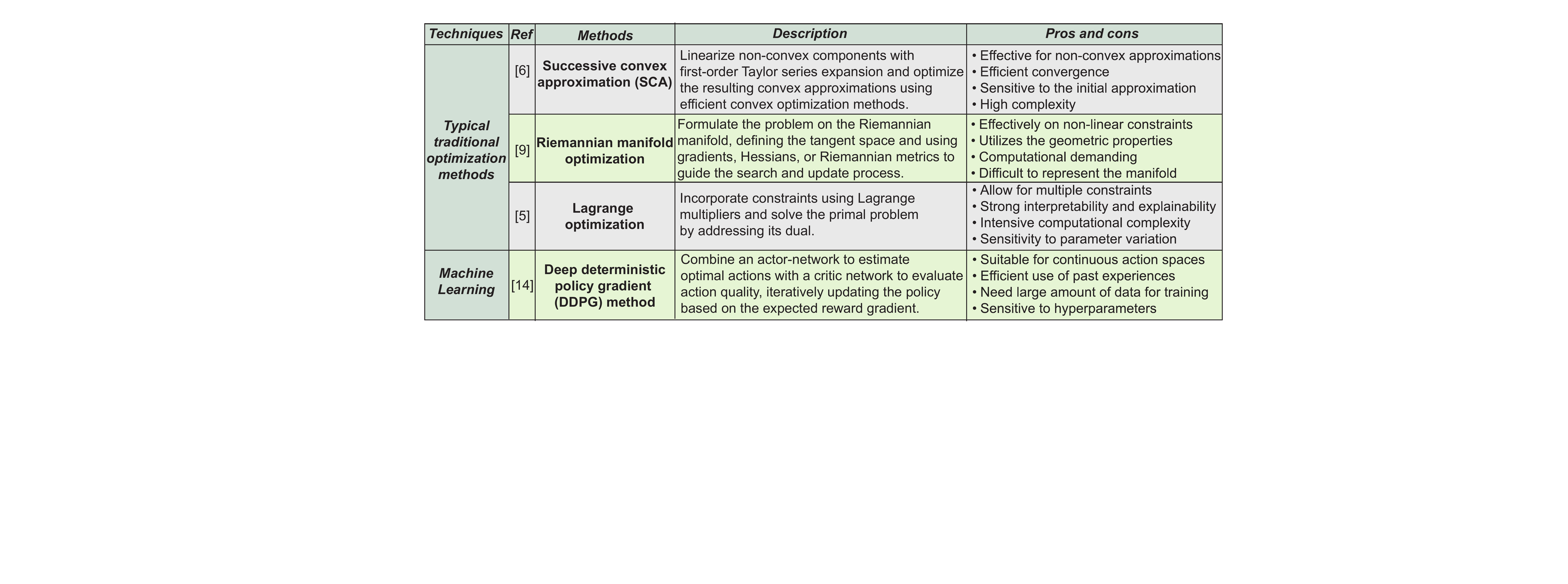}
\vspace{-3mm}
\label{fig:overview}
\end{table*}

{To enhance EE in RIS-aided CF mMIMO systems, our research underscores the importance of joint optimization of AP beamforming and RIS phase shifts. This dual-focus strategy is pivotal for maximizing the reflection performance of the RIS and the beamforming at the APs, which are key variables in the system's optimization. While power control at the AP contributes to improving EE, our focus lies in the iterative optimization of RIS phase shifts and AP beamforming to enhance signal quality and boost EE.} Table \ref{opt} summarizes the characteristics of typical methods and machine learning-based optimization.
\begin{enumerate}
\item \textbf{Successive convex approximation (SCA):} 
{SCA effectively addresses non-convex problems by solving their convex surrogate problems iteratively, suitable for systems needing regular updates with low computational demand and where objectives and constraints are amenable to local convex approximation \cite{lyu2023energy}.}
\item \textbf{Riemannian manifold optimization:} 
{This method leverages Riemannian manifold geometry to efficiently tackle non-linear constraints such as unit-modulus in RIS phase shift design. It is well-suited for systems with the computational capacity to perform complex manifold operations \cite{ma2023cooperative}.}
\item \textbf{Lagrange optimization:} 
{This method employs Lagrange multipliers to address non-convex optimization problems with a mix of equality and inequality constraints, optimized for systems with significant initial conditions through KKT conditions \cite{jin2023energy}.}
\item \textbf{Deep deterministic policy gradient (DDPG) optimization:} {DDPG adopts deep neural networks to train an actor-critic model for continuous control tasks, enabling effective learning and convergence in continuous action spaces, ideal for dynamic environments with changing system parameters \cite{guo2022deep}.}
\end{enumerate}

In RIS-aided CF mMIMO systems, cooperative communication enhances signal transmission through AP-CPU-RIS collaboration. Beam tracking ensures precise beam alignment using methods such as Kalman filters. Joint optimization maximizes energy efficiency by addressing RIS phases and AP beamforming, utilizing traditional methods and machine learning methods. Together, these strategies significantly boost system performance.

\section{Cases study}
This section evaluates and demonstrates the performance of the RIS-induced interference and the beam alignment in RIS-aided CF mMIMO systems. We consider configuring EMI and optimizing beam tracking in a mobile system.

\textbf{Case 1:} 
{In our analysis of the RIS-integrated CF mMIMO system, we specifically addressed the challenges posed by EMI, as depicted in Fig. \ref{case_study}(a). We implemented two power control strategies to tackle these: fractional power control (FPC) and max-min power control. FPC dynamically adjusts transmission power based on UE channel conditions, aiming to mitigate the near-far effect and enhance SE by reducing interference. Conversely, max-min power control is designed to optimize power distribution, focusing on UEs with weaker channels to enhance the system's overall performance.} Furthermore, we compared the system's performance using two combining techniques, i.e., MR combining and LSFD method. {Our research reveals that CF mMIMO systems perform similarly to traditional cellular systems at low energy efficiency levels but outperform them significantly at higher levels, highlighting their potential in high-efficiency environments.} In addition, the findings reveal that the performance of the RIS-aided CF mMIMO system surpasses that of the CF mMIMO system with the max-min SE power control method across all three power allocation methods tested. {Furthermore, the system's performance was favorably influenced by the use of LSFD over MR combining due to its superior signal processing that minimizes interference and enhances SINR, thereby enhancing overall performance. In contrast, despite its simplicity, MR combining fails to effectively handle multi-user interference, which can negatively system efficiency.} {Consequently, both FPC and max-min power control, in conjunction with LSFD, significantly enhance the system's performance compared to full power transmission, with max-min power control and LSFD emerging as the optimal combination \cite{shi2023uplink}.}

\textbf{Case 2:} 
We consider beam tracking in an RIS-aided CF mMIMO system, particularly when the UE is in a mobile state while the AP and RIS are relatively constant. {The CPU initially gathers CSI, which enables APs to mitigate interference through cooperation, unlike traditional cellular systems. Replacing some APs with RIS units enhances beam tracking accuracy and conserves power. The system then utilizes the CSI to adapt the operation of RIS, including RIS location, the number of RIS units, and the number of reflecting elements.} {For efficient beam tracking, we employ algorithms that account for user mobility modeled by a first-order Gauss-Markov model. The EKF updates the estimate and the uncertainty of the current state with the received signal for a moving UE by continuously updating based on mobility data and pilot signals, ensuring precise alignment. This method ensures accurate tracking across diverse UE movement scenarios.} Results in Fig. \ref{case_study} (b) show significant improvements in system performance due to RIS integration. {Larger angles of arrival (AoA) and more RIS elements can improve system performance by covering a broader range of angles of departure (AoD). However, the actual performance gain hinges on the density of RIS elements relative to the AoA variance. If the number of RIS elements does not scale with the increased AoA variance, the system may fail to realize the expected performance improvements, emphasizing the need for a balanced RIS element distribution with the angular spread \cite{zhang2022ris}.}

\section{Conclusions and future directions}
This article explores the potential of RIS-aided CF mMIMO systems for next-generation wireless communication, focusing on power consumption, transmission schemes, and practical deployment. It addresses challenges like RIS-induced interference, beam alignment, and joint optimization, proposing solutions such as cooperative communication, beam tracking, and optimization coupling. Future research directions are also outlined as follows.

\textbf{Generative Artificial Intelligence:} 
{Generative artificial intelligence ({GenAI}) enhances RIS-aided CF mMIMO systems by optimizing beamforming at APs and phase shifts of RIS elements, which improves SE, reduces interference, and maximizes system capacity. For example, generative adversarial networks (GANs) consist of a generator that creates synthetic data and a discriminator that evaluates its authenticity against real data. This process allows GANs to produce realistic channel conditions for training signal processing algorithms. Also, variational autoencoders (VAEs) are generative models that learn to encode and decode data, and can be used to generate new data points similar to the training set. These capabilities allow VAEs to model and generate channel state information for optimization in the RIS-aided CF mMIMO system \cite{guo2022deep}.} {However, integrating GenAI models with existing RIS-aided CF mMIMO systems can be complex, requiring compatibility with legacy systems and infrastructure, as well as ensuring that the AI models do not introduce additional complexity or overhead.}

\textbf{Active RIS}:
{In CF mMIMO systems, the choice between active and passive RIS technologies presents a fair comparison in terms of energy efficiency and system performance. Passive RIS is advantageous for its minimal power consumption and simplicity while facing limitations in high path loss scenarios due to its inability to compensate for signal loss, which can lead to weak signal strength at the receiver, especially after multiple path reflections. On the other hand, active RIS enhances the system's capabilities by integrating active components, such as amplifiers, which actively boost the reflected signals resulting in stronger and more reliable received signals. For example, simultaneously transmitting and reflecting reconfigurable intelligent surface (STAR-RIS) provides flexible signal manipulation by both reflecting and transmitting signals with various designs. Also, multi-functional RIS (MF-RIS) combines reflection, refraction, and amplification, balancing performance and power consumption for advanced applications \cite{zhi2022active}.} {However, active RIS necessitates an efficient allocation of power to its active components, including amplifiers. Indeed, managing this power budget is of paramount importance, as it significantly influences the overall performance of the system.}

%--------------------------------------------------------

\bibliographystyle{IEEEtran}
\bibliography{IEEEabrv,Ref}

\end{document}